\documentclass{article}
\usepackage{graphicx}
\usepackage{amsmath}
\usepackage{amsfonts}
\usepackage{amssymb}
\input epsf

\begin{document}

\title{Study of relativistic bound states in a scalar model using
diagonalization/Monte Carlo methods}
\author{Bu\={g}ra Borasoy$^{a}\thanks{borasoy@physik.tu-muenchen.de}$ and Dean
Lee$^{b}\thanks{dlee@physics.umass.edu}$\\$^{a}${\small Physik Department, Technische Universit\"{a}t M\"{u}nchen,
D-85747 Garching, Germany}\\$^{b}${\small Dept. of Physics, Univ. of Massachusetts, Amherst, MA 01003, U.S.A.}\\\\}
\maketitle
\begin{abstract}
We use a recently proposed diagonalization/Monte Carlo computational scheme to
study relativistic two-body and three-body bound states in $(\phi^{6}-\phi
^{4})_{1+1}$ theory. \ We find that the approach is well-suited for
calculating bound state energies and wavefunctions.
\end{abstract}

\section{Introduction}

The structure and phenomenology of relativistic bound states are of primary
interest to particle and nuclear physics. \ Unfortunately even in the simplest
field theory models our understanding of relativistic bound states is limited
by a number of complications. \ Monte Carlo simulation on the lattice is the
standard tool for determining particle masses. \ One can find the low energy
spectrum from the decay constants of Euclidean correlation functions. \ Though
the procedure is generally successful it becomes more difficult and less
reliable if the state of interest has the same quantum numbers as another
lower energy state. \ Complications also arise for non-confining interactions,
where states just above the continuum threshold can obscure the signal of the
bound state. \ Both of these problems arise even in seemingly simple systems
such as we consider here, namely, two-body and three-body bound states in
$(\phi^{6}-\phi^{4})_{1+1}$.

One remaining complication is of a more operational nature, and it concerns
the interface between computational methods and phenomenology. \ Much of the
interesting and important physics associated with bound states is contained in
subtle details. \ A measurement of gross properties such as the binding energy
is a necessary first step, but one also hopes to gain more descriptive
information like the structure of the wavefunction, its response to various
interactions, or point-by-point comparisons with intuitive non-relativistic models.

These issues were part of the motivation for a recently proposed approach to
computational quantum problems which combines diagonalization and Monte Carlo
techniques \cite{qse, sec}. \ Diagonalization makes it possible to consider
systems with sign oscillations and extract detailed information about
wavefunctions and excited states, while Monte Carlo allows one to handle the
exponential increase in the number of basis states for large volume
systems.\ \ The first half of the method involves finding and diagonalizing
the Hamiltonian restricted to an optimal subspace. \ This subspace is designed
to include the most important basis vectors of the lowest energy eigenstates.
\ Once the most important basis vectors are found and their interactions
treated exactly, Monte Carlo is used to sample the contribution of the
remaining basis vectors. \ In this paper we demonstrate how the new
diagonalization/Monte Carlo scheme is applied to the study of relativistic
bound states.

The essential problem of the lattice Monte Carlo approach to bound states 
near the continuum threshold involves finding the 
inverse Laplace transform of two-point correlation functions at energies 
close to thresholds where the spectral density is large.  This is usually 
done by a non-linear multi-parameter fit of two-point correlators for several 
different operators.  While in principle one might reach arbitrary accuracy 
by adding more parameters to the fit, in practice the presence of even very 
small stochastic error will destabilize the fit and imposes an upper limit 
on the accuracy.  The main advantage of our method, therefore, is not in 
computing speed but in numerical stability.  Our method also provides 
additional information about the wavefunction of the bound state.  These two 
statements are not unrelated.  Our method has greater numerical stability 
because we are implicitly using the information contained in the 
wavefunctions to separate states.
It is also worthwhile noting that
in two dimensions the diagonalization/Monte Carlo approach can be applied
within the framework of discretized light cone and might prove useful since,
aside from zero-mode problems, the exponential growth of the number of basis 
states for large volume systems has so far provided a limit to 
discretized light cone.

The organization of the paper is as follows. \ We begin with a discussion of
the $\phi^{6}-\phi^{4}$ Hamiltonian in $1+1$ dimensions and its decomposition
into momentum modes. \ We then review the basic features of quasi-sparse
eigenvector diagonalization and stochastic error correction. \ In the
following two sections we briefly set aside the numerical discussion and
derive solutions to the non-relativistic versions of the two-body and
three-body bound state problems. \ We then compare our numerical results for
the relativistic bound states energies and wavefunctions with an assortment of
different approximations and results from the literature.

\section{Fock states and the Hamiltonian}

Our starting point is the Hamiltonian density for $\phi^{6}-\phi^{4}$ theory
in $1+1$ dimensions,
\begin{equation}
\mathcal{H}=\text{:}\tfrac{1}{2}\left(  \tfrac{\partial\phi}{\partial
t}\right)  ^{2}+\tfrac{1}{2}\left(  \tfrac{\partial\phi}{\partial x}\right)
^{2}+\tfrac{\mu^{2}}{2}\phi^{2}+\lambda\left(  \phi^{6}-\phi^{4}\right)
\text{:,}%
\end{equation}
where normal ordering is with respect to the mass $\mu$. \ This theory was
introduced in \cite{glimm}\ as a simple model with a\ stable bound state. \ In
$1+1$ dimensions, normal ordering is sufficient to renormalize the theory.

We will consider the theory in a periodic box of length $2L$. \ This allows us
to expand in free particle momentum modes and rewrite the Hamiltonian in terms
of mode creation and annihilation operators \cite{qse, marrero},
\begin{align}
H  &  =%
{\displaystyle\sum_{n}}
\omega_{n}(\mu)a_{n}^{\dagger}a_{n}-\tfrac{\lambda}{8L}%
{\displaystyle\sum_{n_{1}+\cdots+n_{4}=0}}
:\tfrac{\left(  a_{n_{1}}+a_{-n_{1}}^{\dagger}\right)  }{\sqrt{\omega_{n_{1}%
}(\mu)}}\cdots\tfrac{\left(  a_{n_{4}}+a_{-n_{4}}^{\dagger}\right)  }%
{\sqrt{\omega_{n_{4}}(\mu)}}:\label{hamilton}\\
&  +\tfrac{\lambda}{32L^{2}}%
{\displaystyle\sum_{n_{1}+\cdots+n_{6}=0}}
:\tfrac{\left(  a_{n_{1}}+a_{-n_{1}}^{\dagger}\right)  }{\sqrt{\omega_{n_{1}%
}(\mu)}}\cdots\tfrac{\left(  a_{n_{6}}+a_{-n_{6}}^{\dagger}\right)  }%
{\sqrt{\omega_{n_{6}}(\mu)}}:.\nonumber
\end{align}
In (\ref{hamilton}) we have used
\begin{equation}
\omega_{n}(\mu)=\sqrt{\tfrac{n^{2}\pi^{2}}{L^{2}}+\mu^{2}}.
\end{equation}
We can represent any momentum-space Fock state as a string of occupation
numbers, $\left|  o_{-N_{\max}},\cdots,o_{N_{\max}}\right\rangle $, where
\begin{equation}
a_{n}^{\dagger}a_{n}\left|  o_{-N_{\max}},\cdots,o_{N_{\max}}\right\rangle
=o_{n}\left|  o_{-N_{\max}},\cdots,o_{N_{\max}}\right\rangle .
\end{equation}
For the calculations presented here, we set the length of the box to size
$L=2.5\pi\mu^{-1}$ and restrict our attention to momentum modes such that
$\left|  n\right|  \leq N_{\max}$, where $N_{\max}=10$. \ This corresponds
with a momentum cutoff scale of $\Lambda=4\mu.$
A change in the cutoff $N_{\max}$ will lead to adjustments in $\mu$ and
$\lambda$ in order to keep the physics independent of $N_{\max}$.
However, these corrections begin at ${\cal O}(1/ N_{\max}^2)$ as can be checked
by power counting. The dependence on $N_{\max}$ 
has been measured explicitly for 
the bound state data presented in this paper.  It is considerably 
smaller than the dominant error, which is due to higher order 
corrections in the series stochastic error method, see below.

\section{QSE diagonalization}

In order to find approximate eigenvalues and eigenvectors of the Hamiltonian
in (\ref{hamilton}), we use a method called quasi-sparse eigenvector (QSE)
diagonalization. \ QSE diagonalization is an iterative algorithm that finds
the most important basis vectors of the low energy eigenstates for a quantum
Hamiltonian. \ We follow closely the discussion presented in \cite{qse}. \ We
start with a complete basis for which the Hamiltonian matrix $H_{ij}$ is
sparse. \ In our case we use the momentum-space Fock basis. \ The QSE\ method
involves the following iterated steps:

\begin{enumerate}
\item  Select a subset of basis vectors $\left\{  e_{i_{1}},\cdots,e_{i_{n}%
}\right\}  $ and call the corresponding subspace $S$.

\item  Diagonalize $H$ restricted to $S$ and find one eigenvector $v$.

\item  Sort the basis components of $v$ according to their magnitude and
remove the least important basis vectors.

\item  Replace the discarded basis vectors by new basis vectors. \ These are
selected at random from a pool of candidate basis vectors which are connected
to the old basis vectors through non-vanishing matrix elements of $H$.

\item  Redefine $S$ as the subspace spanned by the updated set of basis
vectors and repeat steps 2 through 5.
\end{enumerate}

If the subset of basis vectors is sufficiently large, the exact low energy
eigenvectors will be stable fixed points of the update process. \ This is the
single eigenvector version of the algorithm. \ The multiple eigenvector
version simply involves selecting more than one eigenvector each iteration and
retaining the most important basis vectors for each of the eigenvectors.
One of the most attractive features of 
our method is that we are working directly with the infinite dimensional 
Hilbert space.  There is no restriction on the maximum occupation for any 
mode.  This is a clear advantage over standard diagonalization methods 
such as Lanczos.

In this analysis we work in the rest frame of the bound state and consider
Fock states with total momentum zero. \ Due to $\phi\rightarrow-\phi$
reflection symmetry we can diagonalize the even and odd $\phi$ sectors
separately. \ We use the multiple eigenvector algorithm to obtain the four
lowest eigenstates in the even sector and in the odd sector.\footnote{We have
found only one two-body bound state and one three-body bound state. \ Thus in
a follow-up study one could suffice with fewer states in each sector.} \ For
each QSE iteration, we select $50$ new basis vectors and keep $400$ basis
vectors. \ The final QSE results are used as the starting point for the
stochastic error correction method.

\section{Stochastic error correction}

The QSE method approximates the low energy eigenvalues and eigenvectors of the
Hamiltonian by finding and diagonalizing optimized subspaces $S$. \ This
process however has an intrinsic error due to the omission of basis states not
in $S$. \ The goal of stochastic error correction is to sample the
contribution of these remaining basis vectors. \ In the present investigation
we use the so-called series expansion method up to first order. \ This has
been explained in detail in \cite{sec, lc2000} and therefore we restrict
ourselves to the review of basic formulae.

Let $\lambda_{i}$ be an eigenvalue and $|v_{i}\rangle$ be the corresponding
eigenvector of $H$ restricted to the subspace $S$ (the span of the subset of
basis vectors after step 3 of the QSE algorithm). \ The remaining basis
vectors in the full space not contained in $S$ will be denoted by
$|A_{i}\rangle$, and we are interested in sampling the contribution of these
vectors. \ We assume that $\lambda_{1}$ and $|v_{1}\rangle$ are close to the
corresponding quantities for the full Hamiltonian $H$. \ If we now expand in
powers of
\begin{equation}
\left(  \langle A_{i}|H|A_{i}\rangle-\lambda_{1}\right)  ^{-1}%
\end{equation}
we find that the first order correction to $\lambda_{1}$ is%

\begin{equation}
\delta\lambda_{1}=-\sum_{i}\tfrac{\langle v_{1}|H|A_{i}\rangle\langle
A_{i}|H|v_{1}\rangle}{\langle A_{i}|H|A_{i}\rangle-\lambda_{1}}%
\end{equation}
and the first order correction to $|v_{1}\rangle$ is
\begin{equation}
\left|  \delta v_{1}\right\rangle =-\sum_{i}\tfrac{\left|  A_{i}\right\rangle
\langle A_{i}|H|v_{1}\rangle}{\langle A_{i}|H|A_{i}\rangle-\lambda_{1}}%
+\sum_{j\neq1}\sum_{i}\tfrac{\left|  v_{j}\right\rangle \langle v_{j}%
|H|A_{i}\rangle\langle A_{i}|H|v_{1}\rangle}{(\lambda_{j}-\lambda_{1})(\langle
A_{i}|H|A_{i}\rangle-\lambda_{1})}.
\end{equation}
The sum over all vectors $|A_{i}\rangle$ is too unwieldy to perform exactly.
\ Instead we sample the sum stochastically in the following manner. \ Let
$P(A_{\text{trial}(i)})$ be the probability of selecting $\left|
A_{\text{trial}(i)}\right\rangle $ on the $i^{\text{th}}$ trial. \ Then%

\begin{equation}
\delta\lambda_{1}=-\lim_{N\rightarrow\infty}\frac{1}{N}\sum_{i=1,...,N}%
\tfrac{\langle v_{1}|H|A_{\text{trial}(i)}\rangle\langle A_{\text{trial}%
(i)}|H|v_{1}\rangle}{(\langle A_{\text{trial}(i)}|H|A_{\text{trial}(i)}%
\rangle-\lambda_{1})P(A_{\text{trial}(i)})}%
\end{equation}
and%
\begin{align}
\left\langle w\right.  \left|  \delta v_{1}\right\rangle  &  =-\lim
_{N\rightarrow\infty}\frac{1}{N}\sum_{i}\tfrac{\left\langle w\right.  \left|
A_{\text{trial}(i)}\right\rangle \langle A_{\text{trial}(i)}|H|v_{1}\rangle
}{(\langle A_{\text{trial}(i)}|H|A_{\text{trial}(i)}\rangle-\lambda
_{1})P(A_{\text{trial}(i)})}\\
&  +\lim_{N\rightarrow\infty}\frac{1}{N}\sum_{j\neq1}\sum_{i}\tfrac
{\left\langle w\right.  \left|  v_{j}\right\rangle \langle v_{j}%
|H|A_{\text{trial}(i)}\rangle\langle A_{\text{trial}(i)}|H|v_{1}\rangle
}{(\lambda_{j}-\lambda_{1})(\langle A_{\text{trial}(i)}|H|A_{\text{trial}%
(i)}\rangle-\lambda_{1})P(A_{\text{trial}(i)})}\nonumber
\end{align}
for any fixed vector $\left|  w\right\rangle $.

\section{Two-body bound state}

Before presenting the results of our calculations, we first summarize what we
expect in the weak-coupling non-relativistic limit. \ For the two-body bound
state we consider all two-particle Fock states with zero momentum,%

\begin{equation}
\left|  n,-n\right\rangle =a_{n}^{\dagger}a_{-n}^{\dagger}\left|
0\right\rangle .
\end{equation}
These states satisfy the orthogonality and normalization conditions
\begin{equation}
\left\langle n^{\prime},-n^{\prime}\right|  \left.  n,-n\right\rangle
=\delta_{n,n^{\prime}}+\delta_{n,-n^{\prime}}.
\end{equation}

At lowest order in $\lambda$ the $:\phi^{6}:$ interaction is not relevant to
the two-body bound state problem. \ The effective interaction Hamiltonian is therefore%

\begin{align}
H_{I}  &  =-\lambda:\phi^{4}:\\
&  =-\tfrac{3\lambda}{4L}\sum_{n_{1}+n_{2}+n_{3}+n_{4}=0}\tfrac{a_{-n_{1}%
}^{\dagger}a_{-n_{2}}^{\dagger}a_{n_{3}}a_{n_{4}}}{\left(  \mu^{2}+\frac
{n_{1}^{2}\pi^{2}}{L^{2}}\right)  ^{1/4}\left(  \mu^{2}+\frac{n_{2}^{2}\pi
^{2}}{L^{2}}\right)  ^{1/4}\left(  \mu^{2}+\frac{n_{3}^{2}\pi^{2}}{L^{2}%
}\right)  ^{1/4}\left(  \mu^{2}+\frac{n_{4}^{2}\pi^{2}}{L^{2}}\right)  ^{1/4}%
}.\nonumber
\end{align}
In the non-relativistic limit we can approximate $H_{I}$ as
\begin{equation}
H_{I}=-\tfrac{3\lambda}{4\mu^{2}L}\sum_{n_{1}+n_{2}+n_{3}+n_{4}=0}a_{-n_{1}%
}^{\dagger}a_{-n_{2}}^{\dagger}a_{n_{3}}a_{n_{4}}.
\end{equation}
It is convenient to define the position-space basis vectors
\begin{equation}
\left|  x\right\rangle =\tfrac{1}{\sqrt{2L}}\sum_{n=0,\pm1,...}e^{i\frac{n\pi
x}{L}}\left|  n,-n\right\rangle . \label{x}%
\end{equation}
Since we are considering a periodic box,
\begin{equation}
\left|  x\right\rangle =\left|  x+2L\right\rangle
\end{equation}
and the position-space vectors satisfy
\begin{equation}
\left\langle x^{\prime}\right|  \left.  x\right\rangle =%
{\displaystyle\sum_{j=0,\pm1,...}}
\left[  \delta(x^{\prime}-x+2jL)+\delta(x^{\prime}+x+2jL)\right]  .
\end{equation}

For any function $\Psi(x)$, we can define
\begin{equation}
\left|  \Psi\right\rangle =\int_{-L}^{L}dx\Psi(x)\left|  x\right\rangle .
\end{equation}
If $\left|  \Psi\right\rangle $ is to have unit norm, then we require
\begin{equation}
\int_{-L}^{L}dx\left|  \Psi(x)\right|  ^{2}=\tfrac{1}{2}.
\end{equation}
In our position-space representation of the two-particle subspace, we note
that
\begin{equation}
\sum_{n=0,\pm1,...}a_{n}^{\dagger}a_{n}\left|  x\right\rangle =2\left|
x\right\rangle
\end{equation}
and
\begin{equation}
\sum_{n=0,\pm1,...}\tfrac{n^{2}\pi^{2}}{L^{2}}a_{n}^{\dagger}a_{n}\left|
\Psi\right\rangle =-2\left|  \tfrac{\partial^{2}\Psi}{\partial x^{2}%
}\right\rangle .
\end{equation}

Upon restricting our Hilbert space to the zero-momentum two-particle subspace,
we can simplify the interaction Hamiltonian by the substitution
\begin{equation}
\sum_{n_{1}+n_{2}+n_{3}+n_{4}=0}a_{-n_{1}}^{\dagger}a_{-n_{2}}^{\dagger
}a_{n_{3}}a_{n_{4}}\rightarrow\sum_{n_{1},n_{2}}a_{n_{1}}^{\dagger}a_{-n_{1}%
}^{\dagger}a_{n_{2}}a_{-n_{2}}.
\end{equation}
We observe that
\begin{equation}
\sum_{n_{1},n_{2}}a_{n_{1}}^{\dagger}a_{-n_{1}}^{\dagger}a_{n_{2}}a_{-n_{2}%
}\left|  x\right\rangle =4L\delta(x)\left|  x\right\rangle .
\end{equation}
The full non-relativistic Hamiltonian,
\begin{equation}
H=\sum_{n=0,\pm1,...}\left(  \mu+\tfrac{n^{2}\pi^{2}}{2\mu L^{2}}\right)
a_{n}^{\dagger}a_{n}-\tfrac{3\lambda}{4\mu^{2}L}\sum_{n_{1},n_{2}}a_{n_{1}%
}^{\dagger}a_{-n_{1}}^{\dagger}a_{n_{2}}a_{-n_{2}},
\end{equation}
can now be represented as a Schr\"{o}dinger operator:
\begin{equation}
(H-2\mu)\Psi(x)=\left[  -\tfrac{1}{\mu}\tfrac{\partial^{2}}{\partial x^{2}%
}-\tfrac{3\lambda}{\mu^{2}}\delta(x)\right]  \Psi(x).
\end{equation}
We now consider the eigenvalue equation
\begin{equation}
(H-2\mu)\Psi(x)=E\Psi(x).
\end{equation}
Using the ansatz
\begin{align}
\Psi_{2}(x)  &  \propto e^{-b\left|  x\right|  }+ce^{+b\left|  x\right|  },\\
b  &  >0,
\end{align}
we find
\begin{align}
c  &  =e^{-\tfrac{3\lambda L}{\mu}\tfrac{1+c}{1-c}},\\
b  &  =\tfrac{3\lambda}{2\mu}\tfrac{1+c}{1-c},\\
E  &  =-\tfrac{b^{2}}{\mu}.
\end{align}
When
\begin{equation}
e^{-\tfrac{3\lambda L}{\mu}}<<1,
\end{equation}
we have
\begin{align}
c  &  =e^{-\tfrac{3\lambda L}{\mu}}(1+O(e^{-\tfrac{3\lambda L}{\mu}})),\\
b  &  =\tfrac{3\lambda}{2\mu}(1+O(e^{-\tfrac{3\lambda L}{\mu}})),\\
E  &  =-\tfrac{9\lambda^{2}}{4\mu^{3}}(1+O(e^{-\tfrac{3\lambda L}{\mu}})).
\end{align}
This expression for the binding energy is consistent with the derivation in
\cite{dimock}. \ In the infinite $L$ limit we obtain
\begin{equation}
\Psi_{2}(x)=\sqrt{\tfrac{3\lambda}{4\mu}}e^{-\tfrac{3\lambda\left|  x\right|
}{2\mu}}.
\end{equation}

\section{Three-body bound state}

The analysis for the three-body bound state is slightly more involved. \ In
the non-relativistic weak-coupling limit we restrict our attention to
three-particle states with zero-momentum,%

\begin{equation}
\left|  n_{1},n_{2},-n_{1}-n_{2}\right\rangle =a_{n_{1}}^{\dagger}a_{n_{2}%
}^{\dagger}a_{-n_{1}-n_{2}}^{\dagger}\left|  0\right\rangle .
\end{equation}
We again construct position-space basis states
\begin{equation}
\left|  x_{1},x_{2}\right\rangle =\tfrac{\sqrt{3}}{2L}\sum_{n_{1},n_{2}%
=0,\pm1,...}e^{i\frac{n_{1}\pi x_{1}}{L}}e^{i\frac{n_{2}\pi x_{2}}{L}%
}e^{i\frac{(-n_{1}-n_{2})\pi(-x_{1}-x_{2})}{L}}\left|  n_{1},n_{2}%
,-n_{1}-n_{2}\right\rangle ,
\end{equation}
which satisfy the orthogonality and normalization conditions%
\begin{equation}
\left\langle x_{1}^{\prime},x_{2}^{\prime}\right|  \left.  x_{1}%
,x_{2}\right\rangle =%
{\displaystyle\sum_{j,k=0,\pm1,...}}
\left[
\begin{array}
[c]{c}%
\delta(x_{1}^{\prime}-x_{1}+(\frac{2j}{3}+2k)L)\delta(x_{2}^{\prime}%
-x_{2}+\frac{2j}{3}L)\\
+\left\{  x_{1}\leftrightarrow x_{2}\leftrightarrow-x_{1}-x_{2}\right\}
\end{array}
\right]  .
\end{equation}
The periodicity of our system implies
\begin{align}
\left|  x_{1},x_{2}\right\rangle  &  =\left|  x_{1}+\tfrac{2}{3}L,x_{2}%
+\tfrac{2}{3}L\right\rangle \label{period}\\
&  =\left|  x_{1}+2L,x_{2}\right\rangle =\left|  x_{1},x_{2}+2L\right\rangle
.\nonumber
\end{align}
For any function $\Psi(x_{1},x_{2})$, we define%
\begin{equation}
\left|  \Psi\right\rangle =\int_{R}dx_{1}dx_{2}d\Psi(x_{1},x_{2})\left|
x_{1},x_{2}\right\rangle ,
\end{equation}
where $R$ is the region%
\begin{equation}
R=\left\{
\begin{array}
[c]{c}%
-L<x_{1}\leq L\\
-\tfrac{1}{3}L<x_{2}\leq\tfrac{1}{3}L.
\end{array}
\right.
\end{equation}
If $\left|  \Psi\right\rangle $ is to have unit norm then we must have%
\begin{equation}
\int_{R}dx_{1}dx_{2}\left|  \Psi(x_{1},x_{2})\right|  ^{2}=\tfrac{1}{6}.
\end{equation}

We note that
\begin{equation}
\sum_{n=0,\pm1,...}\tfrac{n^{2}\pi^{2}}{L^{2}}a_{n}^{\dagger}a_{n}\left|
\Psi\right\rangle =-\tfrac{2}{3}\left|  \tfrac{\partial^{2}\Psi}{\partial
x_{1}^{2}}+\tfrac{\partial^{2}\Psi}{\partial x_{2}^{2}}-\tfrac{\partial
^{2}\Psi}{\partial x_{1}\partial x_{2}}\right\rangle .
\end{equation}
With a little work we also find%
\begin{align}
&  \sum_{n_{1}+n_{2}+n_{3}+n_{4}=0}a_{-n_{1}}^{\dagger}a_{-n_{2}}^{\dagger
}a_{n_{3}}a_{n_{4}}\left|  x_{1},x_{2}\right\rangle \\
&  =4L\sum_{j}\left[  \delta(x_{1}+2x_{2}+2jL)+\delta(2x_{1}+x_{2}%
+2jL)+\delta(x_{1}-x_{2}+2jL)\right]  \left|  x_{1},x_{2}\right\rangle
.\nonumber
\end{align}
For the moment we will neglect the $:\phi^{6}:$ term. \ Later we will show
that this term produces only a higher order correction in $\lambda$. \ The
effective non-relativistic Hamiltonian, \
\begin{equation}
H=\sum_{n=0,\pm1,...}\left(  \mu+\tfrac{n^{2}\pi^{2}}{2\mu L^{2}}\right)
a_{n}^{\dagger}a_{n}-\tfrac{3\lambda}{4\mu^{2}L}\sum_{n_{1}+n_{2}+n_{3}%
+n_{4}=0}a_{-n_{1}}^{\dagger}a_{-n_{2}}^{\dagger}a_{n_{3}}a_{n_{4}},
\end{equation}
can now be represented as a Schr\"{o}dinger operator:%
\begin{align}
(H-3\mu)\Psi &  =-\tfrac{1}{3\mu}\left(  \tfrac{\partial^{2}\Psi}{\partial
x_{1}^{2}}+\tfrac{\partial^{2}\Psi}{\partial x_{2}^{2}}-\tfrac{\partial
^{2}\Psi}{\partial x_{1}\partial x_{2}}\right) \\
&  -\tfrac{3\lambda}{\mu^{2}}\sum_{j}\left[
\begin{array}
[c]{c}%
\delta(x_{1}+2x_{2}+2jL)+\delta(2x_{1}+x_{2}+2jL)\\
+\delta(x_{1}-x_{2}+2jL)
\end{array}
\right]  \Psi.\nonumber
\end{align}
In the infinite $L$ limit we can write the energy eigenvalue equation as%
\begin{equation}
-\tfrac{1}{3\mu}\left(  \tfrac{\partial^{2}\Psi}{\partial x_{1}^{2}}%
+\tfrac{\partial^{2}\Psi}{\partial x_{2}^{2}}-\tfrac{\partial^{2}\Psi
}{\partial x_{1}\partial x_{2}}\right)  -\tfrac{3\lambda}{\mu^{2}}\left[
\begin{array}
[c]{c}%
\delta(x_{1}+2x_{2})+\delta(2x_{1}+x_{2})\\
+\delta(x_{1}-x_{2})
\end{array}
\right]  \Psi=E\Psi.
\end{equation}
Using the ansatz%
\begin{equation}
\Psi_{3}\propto e^{-b(\left|  x_{1}+2x_{2}\right|  +\left|  2x_{1}%
+x_{2}\right|  +\left|  x_{1}-x_{2}\right|  )},
\end{equation}
we find%
\begin{align}
b  &  =\tfrac{3\lambda}{2\mu},\\
E  &  =-\tfrac{9\lambda^{2}}{\mu^{3}}.
\end{align}
>From the normalization condition,%
\begin{equation}
\Psi_{3}=\tfrac{\sqrt{3}\lambda}{\mu}e^{-\tfrac{3\lambda}{2\mu}(\left|
x_{1}+2x_{2}\right|  +\left|  2x_{1}+x_{2}\right|  +\left|  x_{1}%
-x_{2}\right|  )}. \label{wave}%
\end{equation}

We now return to the $:\phi^{6}:$ term. \ This term produces an interaction
proportional to \
\begin{equation}
\tfrac{\lambda}{\mu^{3}L^{2}}\delta(x_{1})\delta(x_{2}).
\end{equation}
>From (\ref{wave}) we see that%
\begin{equation}
\left\langle \Psi_{3}\right|  \delta(x_{1})\delta(x_{2})\left|  \Psi
_{3}\right\rangle =O(\lambda^{2}),
\end{equation}
and therefore the $:\phi^{6}:$ interaction produces an effect of order
$O(\lambda^{3})$. \ We expect however that the repulsive delta function will
suppress $\Psi_{3}$ near $x_{1}=x_{2}=0$.

\section{Results}

In Table 1 we show for several values of $\lambda$
the mass of the one-particle state, $m_{1}$, and the
two-body bound state, $m_{2}$, which is the relativistic state that is
continously connected to the non-relativistic two-body bound state. \ For
comparison we show the data for $m_{1}$ and $m_{2}$ from the lattice
diagonalization study of \cite{barnes} (BD), non-relativistic Schr\"{o}dinger
results derived above for infinite $L$ (S) and $L=2.5\pi\mu^{-1}$ (L), and
Gaussian effective potential results from \cite{darewych, stevenson} (GEP).
\ The remaining column of data comes from an exact diagonalization of the
Hamiltonian restricted to two-particle Fock states (F). \ In order to obtain
the ratio $m_{2}/m_{1}$ in this approximation we take $m_{1}=\mu
.$\footnote{This is at least consistent since in the two-particle Fock
approximation with normal-ordered interactions, the constituents of the bound
state have no self-energy contributions.}
\[
\overset{\text{Table 1}}{%
\begin{tabular}
[c]{|l|l|l|l|l|l|l|l|l|}\hline
$\lambda/\mu^{2}$ & $m_{1}/\mu$ & BD & $m_{2}/m_{1}$ & S & L & F & BD &
GEP\\\hline
$0.1$ & $0.995(1)$ & $0.994$ & $1.970(2)$ & $1.978$ & $1.967$ & $1.975$ &
$1.916$ & $1.982$\\\hline
$0.2$ & $0.975(5)$ & $0.972$ & $1.92(1)$ & $1.910$ & $1.907$ & $1.934$ &
$1.821$ & $1.938$\\\hline
$0.3$ & $0.94(1)$ & $0.928$ & $1.85(1)$ & $1.798$ & $1.797$ & $1.877$ &
$1.718$ & $1.878$\\\hline
$0.4$ & $0.88(2)$ & $0.850$ & $1.76(4)$ & $1.640$ & $1.640$ & $1.808$ &
$1.613$ & $1.807$\\\hline
$0.5$ & $0.77(4)$ & $0.720$ & $1.66(8)$ & $1.438$ & $1.438$ & $1.730$ &
$1.520$ & $1.728$\\\hline
$0.6$ & $0.61(4)$ &  & $1.71(10)$ & $1.190$ & $1.190$ & $1.646$ &  &
$1.642$\\\hline
\end{tabular}
}%
\]
\ The error bars in Table 1 and throughout our discussion include estimated
errors from Monte Carlo errors, higher order contributions in the stochastic
error correction series expansion, and residual dependence on the ultraviolet
cutoff $\Lambda.$ \ Of these errors we find the higher order contributions in
the expansion to be the largest source of error. \ Any follow-up study of this
system should push the calculation to higher order or make use of the
stochastic Lanczos method introduced in \cite{sec}.

The results from \cite{barnes} (BD) were carried out on a much smaller system.
\ While their results for $m_{1}$ appear in agreement with ours, they have
noted that their calculation for $m_{2}$ is considerably affected by the small
system size. \ At small coupling at least our own finite size effects appear
to be small, as is suggested by the agreement of the Schr\"{o}dinger results S
and L. \ For this range of couplings the GEP predictions are equivalent to the
two-particle Fock state approximation (F) in the limit of infinite $L$ and
$\Lambda$. \ This is because at these couplings $\bar{\Omega}=m_{1}=\mu,$
where $\bar{\Omega}$ is the optimal GEP variational mass parameter as defined
in \cite{darewych}$.$

As expected, our results agree with the Schr\"{o}dinger results at small
coupling. \ They also appear reasonably close to the two-particle Fock
approximation, F, for all values of coupling. \ In our next calculation we
measure how much of the bound state does in fact reside in the two-particle
sector. \ Let $\left|  m_{2}\right\rangle $ be the normalized bound state and
$P_{2}$ be the projection operator onto the two-particle subspace. \ We define
$N_{2}$ to be the norm of the two-particle projection \
\begin{equation}
N_{2}=\left\|  P_{2}\left|  m_{2}\right\rangle \right\|  .
\end{equation}
In Table 2 we show $N_{2}$ for the different couplings.\ \
\[
\overset{\text{Table 2}}{%
\begin{tabular}
[c]{|l|l|l|l|l|l|l|}\hline
$\lambda/\mu^{2}$ & $0.1$ & $0.2$ & $0.3$ & $0.4$ & $0.5$ & $0.6$\\\hline
$N_{2}$ & $0.9977(1)$ & $0.989(1)$ & $0.971(2)$ & $0.936(8)$ & $0.87(2)$ &
$0.79(2)$\\\hline
\end{tabular}
}%
\]
We can pursue the question of the bound state wavefunction further. \ Let%
\begin{equation}
\Psi_{2}(x)=\tfrac{1}{N_{2}}\left\langle x\right|  \left.  m_{2}\right\rangle
,
\end{equation}
where $\left|  x\right\rangle $ is the two-particle position state defined in
($\ref{x}$)$.$ \ In Figures 1-3 we compare the wavefunctions $\Psi_{2}$ with
the corresponding approximate wavefunctions $\Psi_{2}^{\text{S}}$, $\Psi
_{2}^{\text{L}}$, $\Psi_{2}^{\text{F}}$ for coupling values $\lambda
=0.1\mu^{2},0.2\mu^{2},0.4\mu^{2}$.

In our analysis we did not consider bound states beyond $\lambda=0.6\mu^{2}$.
\ This is because our diagonalization/Monte Carlo results indicate a first
order phase transition near $\lambda=0.73\mu^{2}$, where $\phi$ develops a
vacuum expectation value \
\begin{equation}
\left\langle \phi\right\rangle =1.02(2).
\end{equation}
As a matter of fact, there appears to be no bound state in the broken
symmetry phase, and the two-body 
and three-body bound states in the symmetric phase are stable all 
the way up to the transition point.
The phase transition can be seen quite clearly in the behavior of the energy
levels. \ In Table 3 we show the lowest two energy levels $E_{0}$ and $E_{2}$
in the even $\phi$ sector. \ We see what appears to be a metastable vacuum at
energy $E_{2}$ for $\lambda<0.73\mu^{2}$ becoming the true vacuum at energy
$E_{0}$ for $\lambda>0.73\mu^{2}$.\footnote{Since we are at finite volume the
levels $E_{0}$ and $E_{2}$ come close but never actually become degenerate.}
\
\[
\overset{\text{Table 3}}{%
\begin{tabular}
[c]{|l|l|l|}\hline
$\lambda/\mu^{2}$ & $E_{0}$ & $E_{2}$\\\hline
$0.70$ & $-0.37(5)$ & $0.09(10)$\\\hline
$0.71$ & $-0.41(5)$ & $-0.15(15)$\\\hline
$0.72$ & $-0.40(5)$ & $-0.29(20)$\\\hline
$0.73$ & $-0.49(20)$ & $-0.42(5)$\\\hline
$0.74$ & $-0.70(20)$ & $-0.42(5)$\\\hline
\end{tabular}
}%
\]

In Table 4 we show for several values of $\lambda$ 
the mass of the three-body bound state, $m_{3}$, 
which is the relativistic state that is
continously connected to the non-relativistic three-body bound state. \ We 
were not able to find any previous studies
of this state in the literature. \ For comparison we show data from the
Schr\"{o}dinger equation at infinite $L$ (S) and two different Fock space
approximations, F$_{1}$ and F$_{2}$. \ F$_{1}$ corresponds with an exact
diagonalization of the Hamiltonian restricted to three-particle Fock states,
and F$_{2}$ corresponds to the same thing, except with the $:\phi^{6}:$
interaction turned off.\footnote{This is an artificial approximation.
\ Without the restriction to three-particle Fock states, turning off the
$:\phi^{6}:$ interaction would make the theory unbounded below.} \ To obtain
the ratio $m_{3}/m_{1}$ in the two Fock approximations we again take
$m_{1}=\mu.$%
\[
\overset{\text{Table 4}}{%
\begin{tabular}
[c]{|l|l|l|l|l|}\hline
$\lambda/\mu^{2}$ & $m_{3}/m_{1}$ & S & F$_{1}$ & F$_{2}$\\\hline
$0.1$ & $2.931(2)$ & $2.910$ & $2.940$ & $2.919$\\\hline
$0.2$ & $2.83(1)$ & $2.640$ & $2.862$ & $2.770$\\\hline
$0.3$ & $2.72(2)$ & $2.190$ & $2.769$ & $2.568$\\\hline
$0.4$ & $2.59(6)$ & $1.560$ & $2.667$ & $2.336$\\\hline
$0.5$ & $2.54(8)$ & $0.75$ & $2.558$ & $2.084$\\\hline
$0.6$ & $2.62(15)$ & $<0$ & $2.444$ & $1.818$\\\hline
\end{tabular}
}%
\]
We notice that the Schr\"{o}dinger approximation\ (S) is not quite as accurate
for given $\lambda$ as it was for the two-body bound state. \ Nevertheless the
Fock approximation F$_{1}$ still appears to follow the actual data relatively
well. \ We observe that $m_{3}$ is not only below the $3m_{1}$ continuum
threshold but also below the $m_{2}+m_{1}$ threshold. \ The corresponding
state therefore has a sensible interpretation as a stable particle.

In analogy with before, we let $\left|  m_{3}\right\rangle $ be the normalized
three-body bound state and $P_{3}$ be the projection operator onto the
three-particle subspace. \ We define $N_{3}$ to be the norm of the
three-particle projection. \
\begin{equation}
N_{3}=\left\|  P_{3}\left|  m_{3}\right\rangle \right\|  .
\end{equation}
In Table 5 we show $N_{3}$ for different couplings. \ %

\[
\overset{\text{Table 5}}{%
\begin{tabular}
[c]{|l|l|l|l|l|l|l|}\hline
$\lambda/\mu^{2}$ & $0.1$ & $0.2$ & $0.3$ & $0.4$ & $0.5$ & $0.6$\\\hline
$N_{3}$ & $0.9966(2)$ & $0.981(2)$ & $0.950(5)$ & $0.89(1)$ & $0.80(2)$ &
$0.66(4)$\\\hline
\end{tabular}
}%
\]
Let%
\begin{equation}
\Psi_{3}(x_{1},x_{2})=\tfrac{1}{N_{3}}\left\langle x_{1},x_{2}\right|  \left.
m_{3}\right\rangle .
\end{equation}
In Figures 4-11 we compare the wavefunction $\Psi_{3}$ with the corresponding
approximate wavefunctions $\Psi_{3}^{\text{S}}$, $\Psi_{3}^{\text{F}_{1}}$,
$\Psi_{3}^{\text{F}_{2}}$ for $\lambda=0.1\mu^{2}$ and $0.4\mu^{2}$. \ For
visual clarity we plot the wavefunctions over three copies of the fundamental
region $R$. \ We use the variables
\begin{align}
y_{1}  &  =\tfrac{3}{2L}x_{1},\\
y_{2}  &  =\tfrac{\sqrt{3}}{2L}\left(  x_{1}+2x_{2}\right)  ,
\end{align}
to make clear the hexagon symmetry. \ To guide the eye we have drawn dotted
lines along the points%
\begin{align}
x_{1}  &  =x_{2},\\
x_{1}  &  =-x_{1}-x_{2},\\
x_{2}  &  =-x_{1}-x_{2}.
\end{align}
These corresponds with configurations where two of the three constituents lie
at the same point.

We notice that for both $\Psi_{3}$ and $\Psi_{3}^{\text{F}_{1}},$ the
inclusion of the repulsive $:\phi^{6}:$ interaction produces a marked
suppression of the wavefunction near the origin. \ By independently changing
the coefficients of $:\phi^{6}:$ and $:\phi^{4}:$, we have a simple laboratory
to study the competition between two-body and three-body forces in three-body
bound states, a topic of some relevance and recent interest in the triton
system. \cite{bedaque}

\section{Summary}

We have considered the relativistic bound states of $\phi^{6}-\phi^{4}$ theory
in $1+1$ dimensions. \ Using a recently proposed diagonalization/Monte Carlo
computational scheme, we calculated the two-body and three-body bound state
energies and wavefunctions. \ The initial diagonalization was performed using
the quasi-sparse eigenvector method, and the subsequent stochastic error
correction was calculated to first-order using the series method. \ We made
detailed comparisons of our numerical results with the available literature
and several approximations, including non-relativistic weak-coupling solutions
and various truncated Fock space approximations.

Applications to more complicated higher dimensional theories can be done in
the same manner but require larger QSE subspaces and going beyond the
first-order term in the series method or use of the stochastic Lanczos method.
\ But even from the simple examples presented here, the potential for studying
relativistic bound states with diagonalization/Monte Carlo methods is apparent.

\paragraph*{Acknowledgments}

\bigskip\textsl{Financial support provided by the DFG and NSF.}

\begin{figure}[t]
\begin{center}
\epsfxsize=23pc \epsfbox{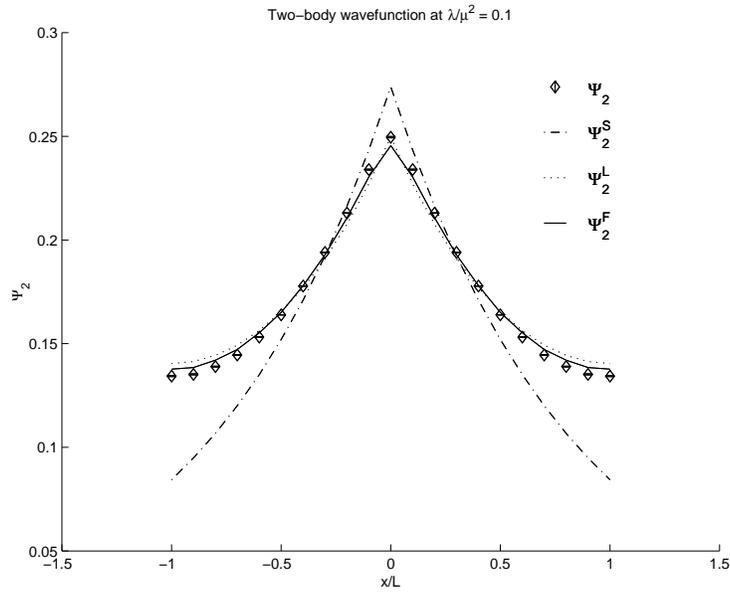}
\end{center}
\caption{Two-body wavefunction $\Psi_{2}$ and approximate wavefunctions
$\Psi_{2}^{\text{S}},\Psi_{2}^{\text{L}},\Psi_{2}^{\text{F}}$ at
$\lambda=0.1\mu^{2}.$}%
\end{figure}\begin{figure}[tt]
\begin{center}
\epsfxsize=21pc \epsfbox{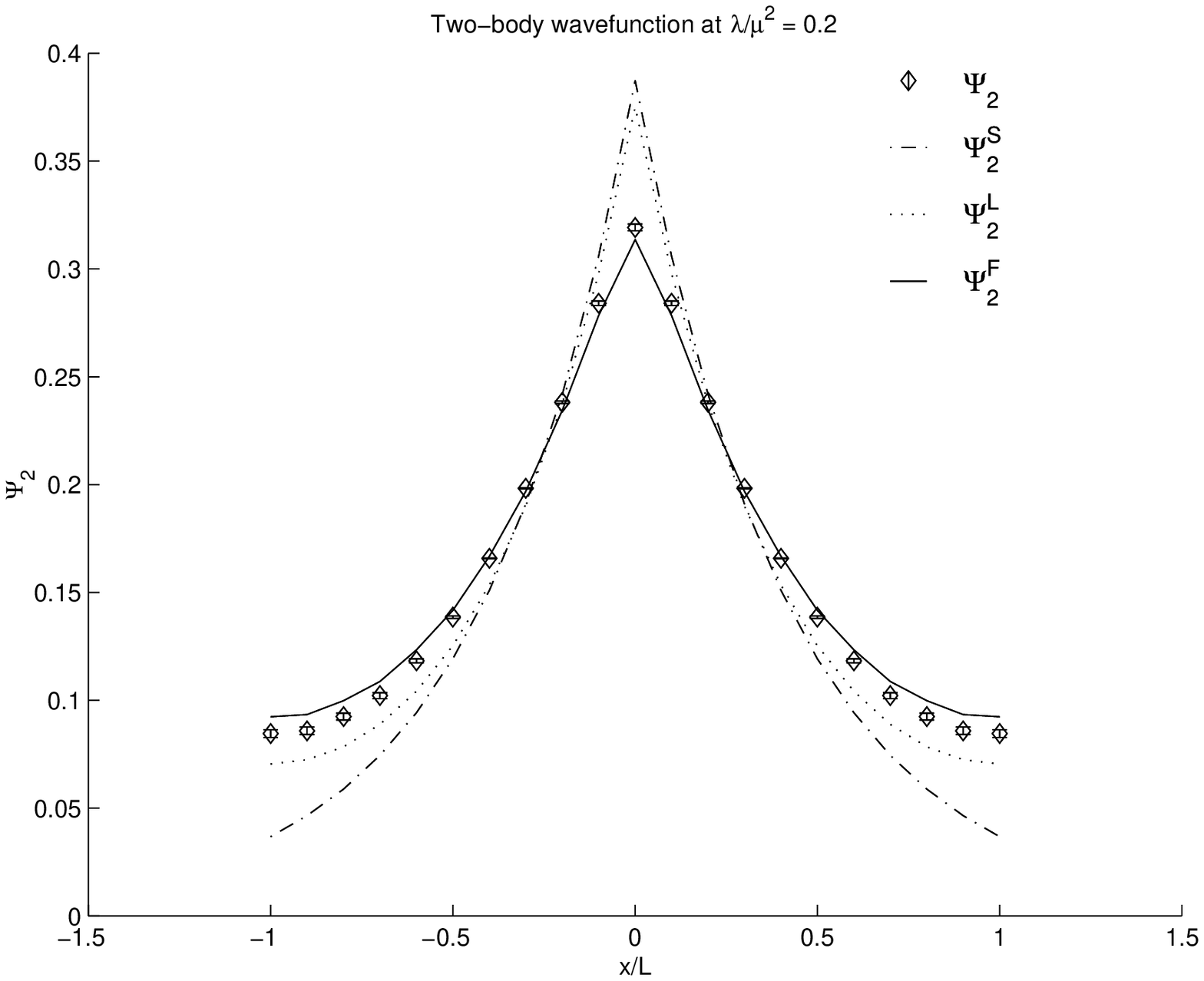}
\end{center}
\caption{Two-body wavefunction $\Psi_{2}$ and approximate wavefunctions
$\Psi_{2}^{\text{S}},\Psi_{2}^{\text{L}},\Psi_{2}^{\text{F}}$ at
$\lambda=0.2\mu^{2}.$}%
\end{figure}\begin{figure}[ttt]
\begin{center}
\epsfxsize=21pc \epsfbox{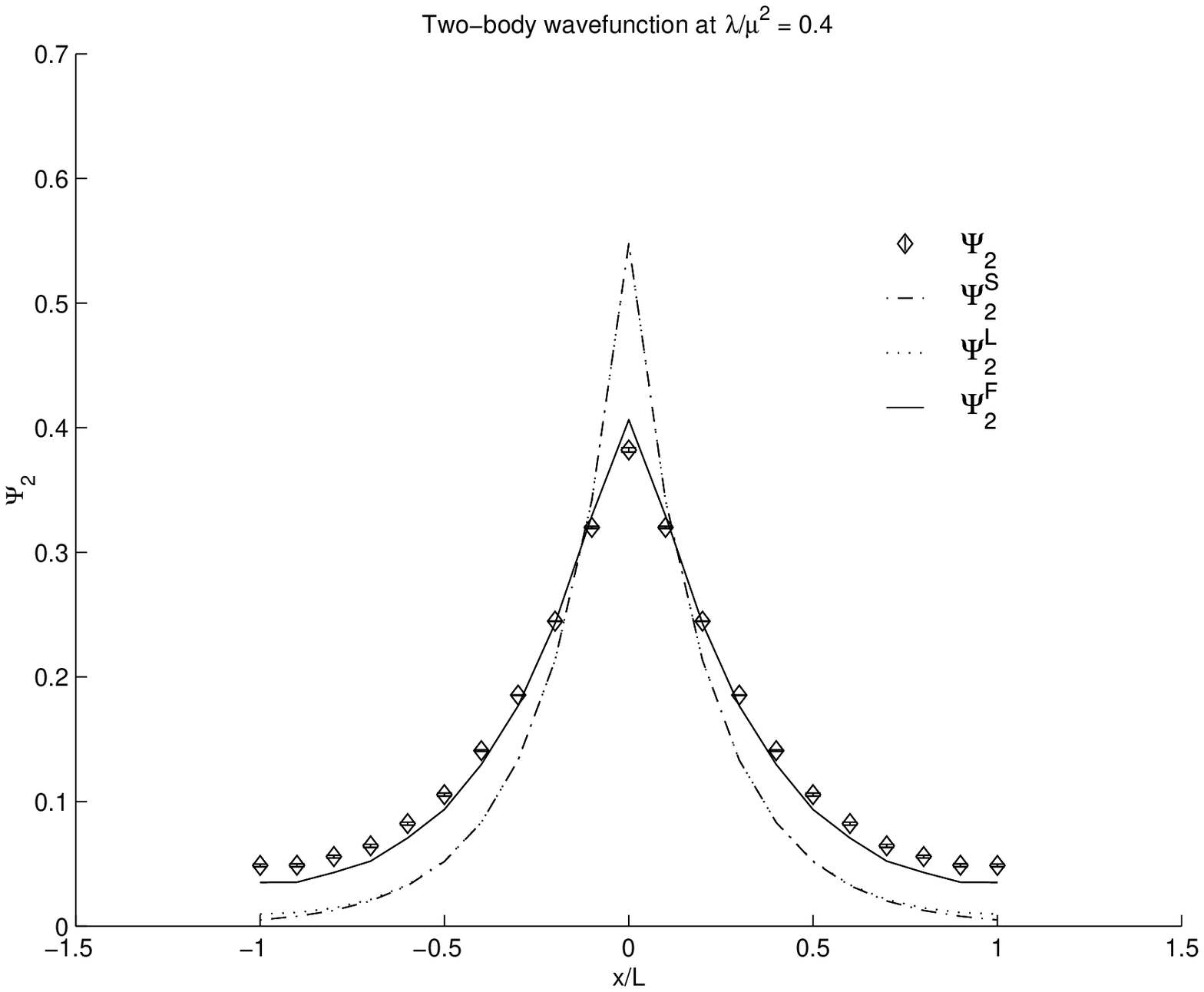}
\end{center}
\caption{Two-body wavefunction $\Psi_{2}$ and approximate wavefunctions
$\Psi_{2}^{\text{S}},\Psi_{2}^{\text{L}},\Psi_{2}^{\text{F}}$ at
$\lambda=0.4\mu^{2}.$}%
\end{figure}\begin{figure}[tttt]
\begin{center}
\epsfxsize=21pc \epsfbox{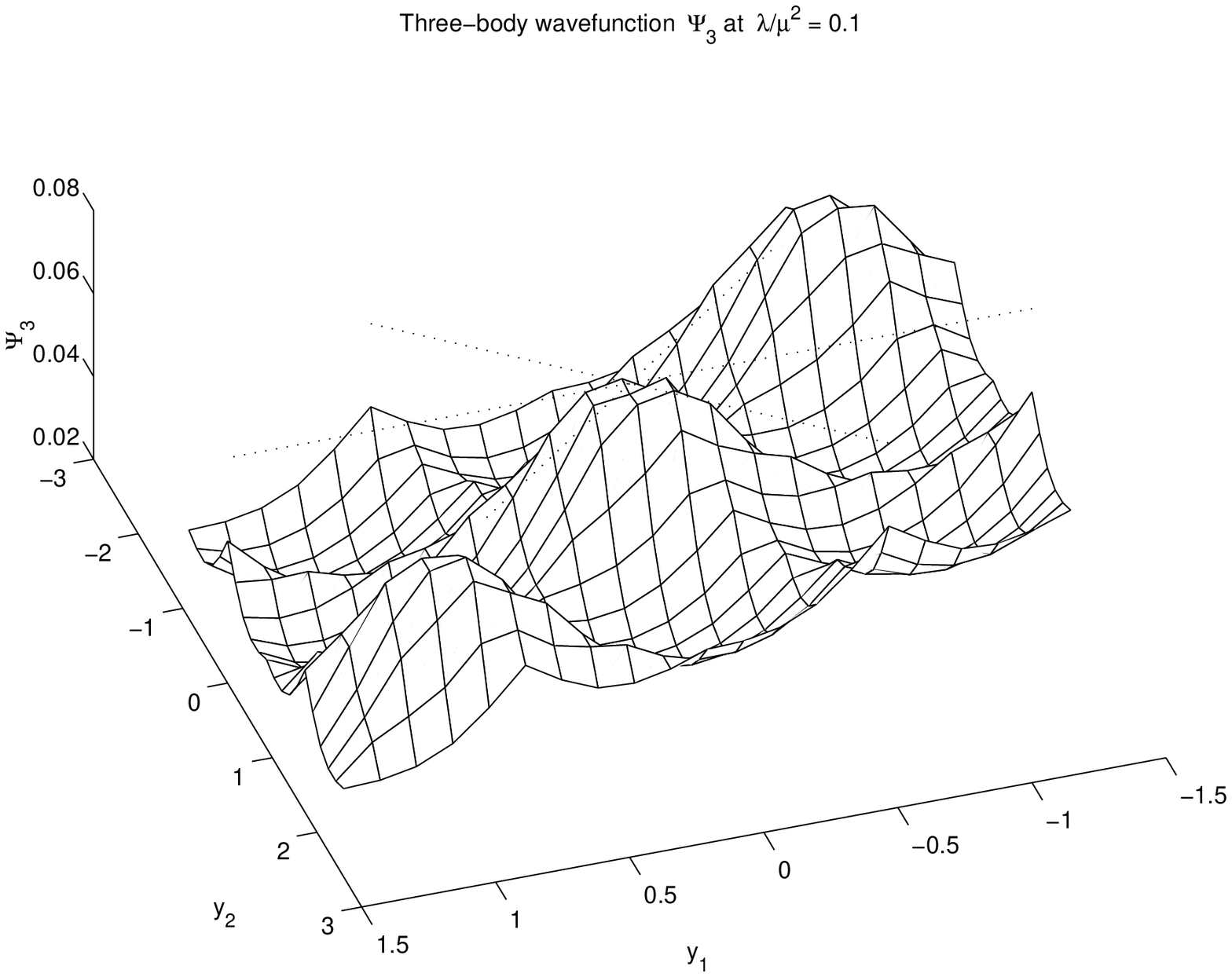}
\end{center}
\caption{Three-body wavefunction $\Psi_{3}$ at $\lambda=0.1\mu^{2}.$}%
\end{figure}\begin{figure}[ttttt]
\begin{center}
\epsfxsize=23pc \epsfbox{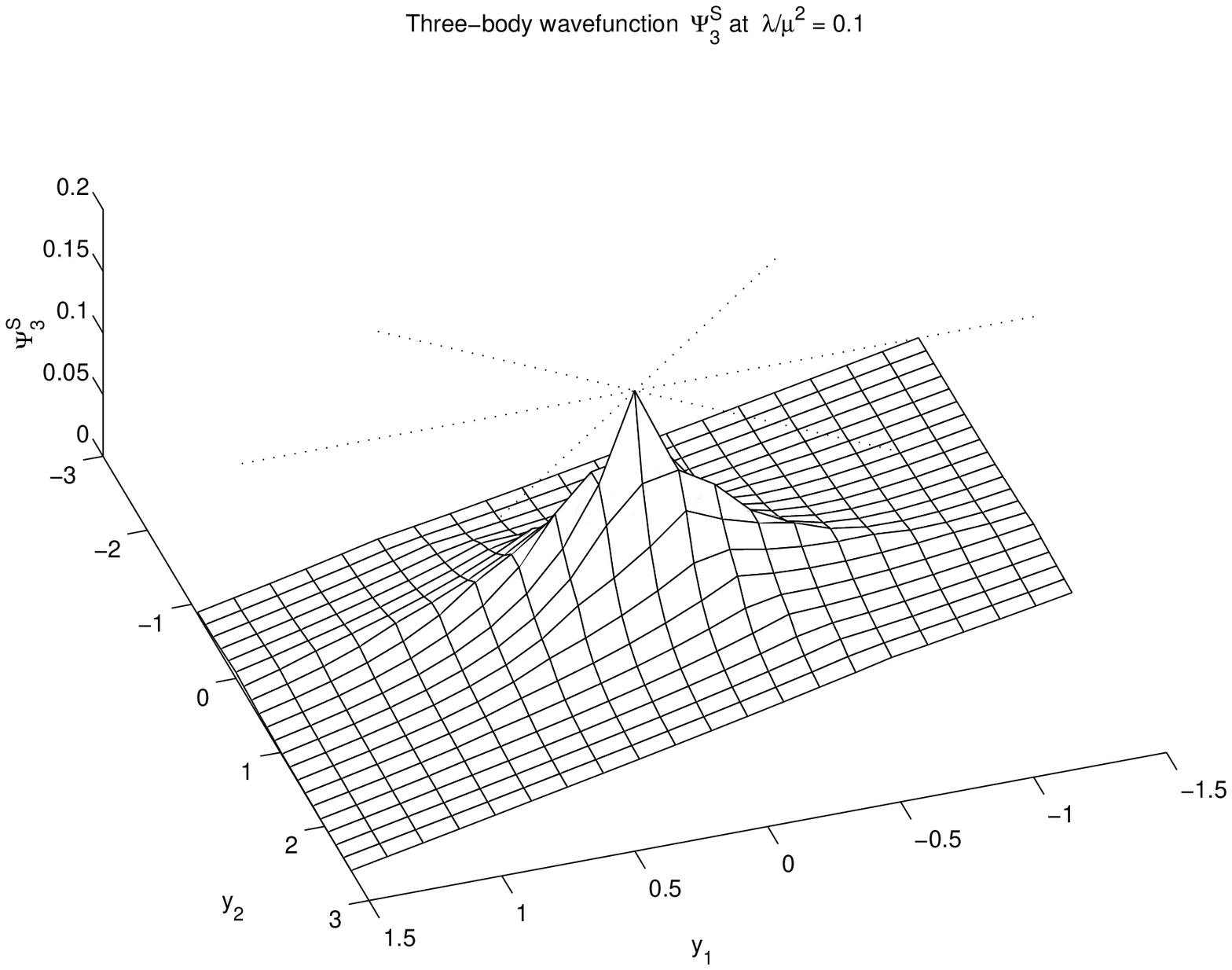}
\end{center}
\caption{Approximate three-body wavefunction $\Psi_{3}^{\text{S}}$ at
$\lambda=0.1\mu^{2}.$}%
\end{figure}\begin{figure}[tttttt]
\begin{center}
\epsfxsize=23pc \epsfbox{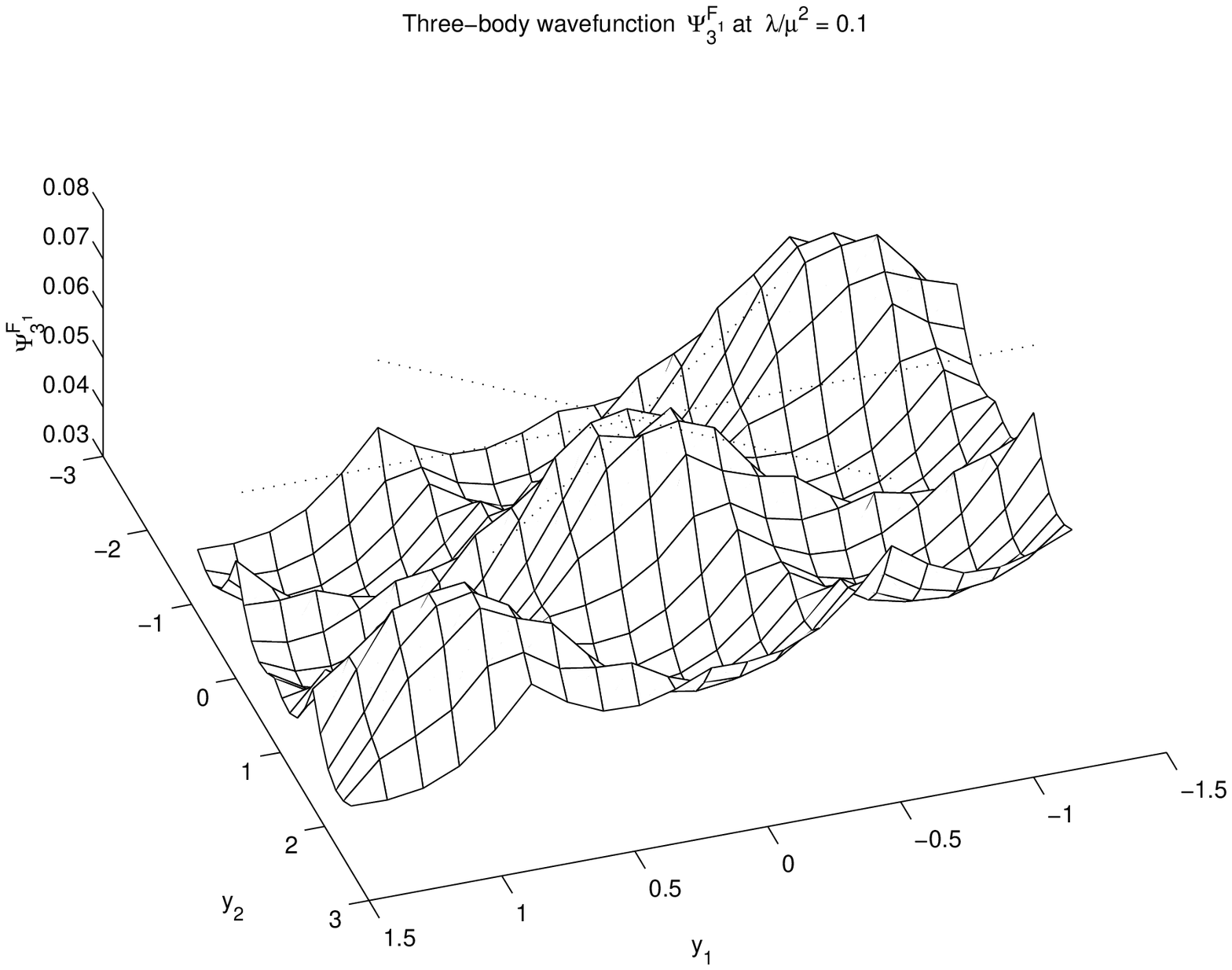}
\end{center}
\caption{Approximate three-body wavefunction $\Psi_{3}^{\text{F}_{1}}$ at
$\lambda=0.1\mu^{2}.$}%
\end{figure}\begin{figure}[ttttttt]
\begin{center}
\epsfxsize=23pc \epsfbox{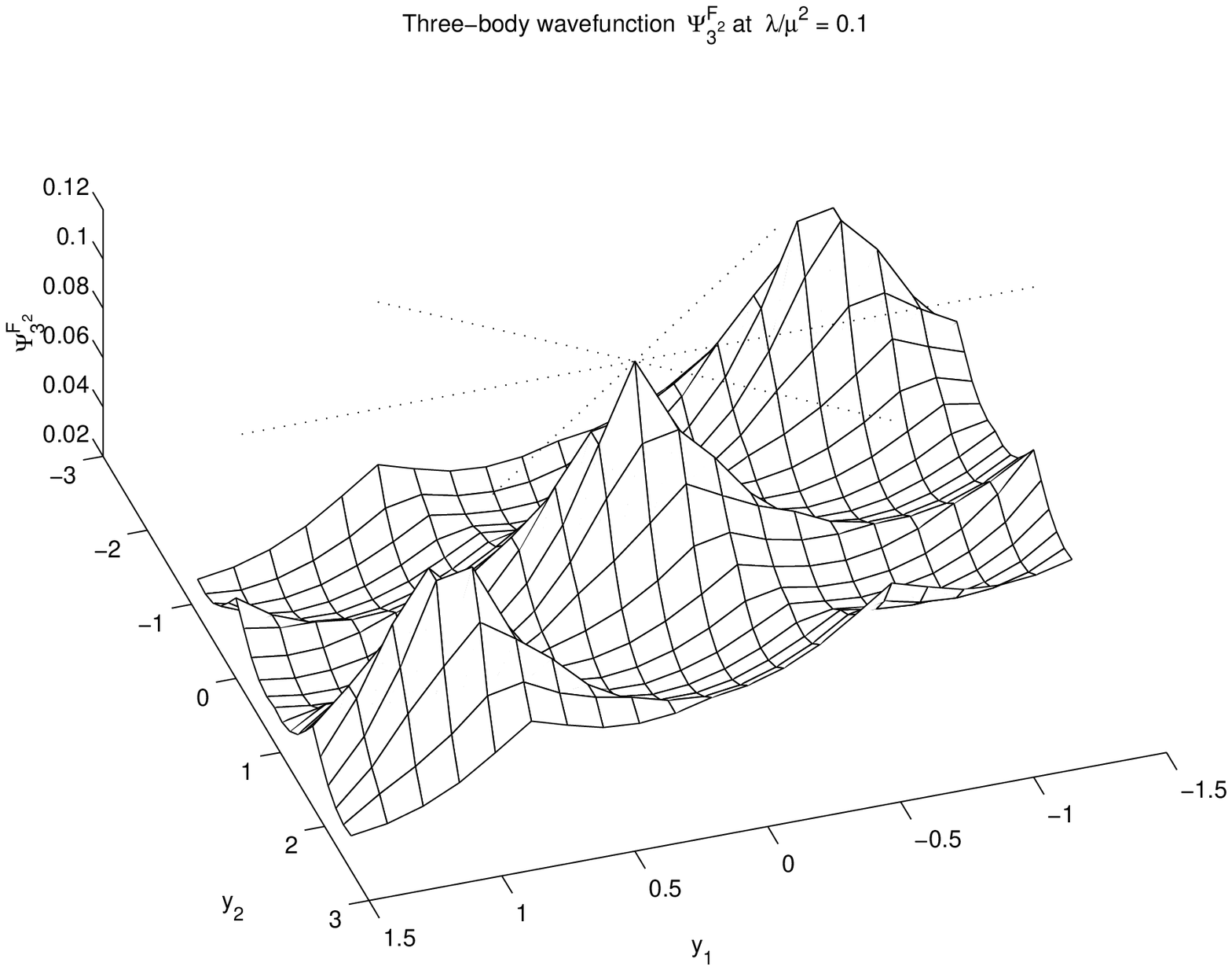}
\end{center}
\caption{Approximate three-body wavefunction $\Psi_{3}^{\text{F}_{2}}$ at
$\lambda=0.1\mu^{2}.$}%
\end{figure}\begin{figure}[tttttttt]
\begin{center}
\epsfxsize=23pc \epsfbox{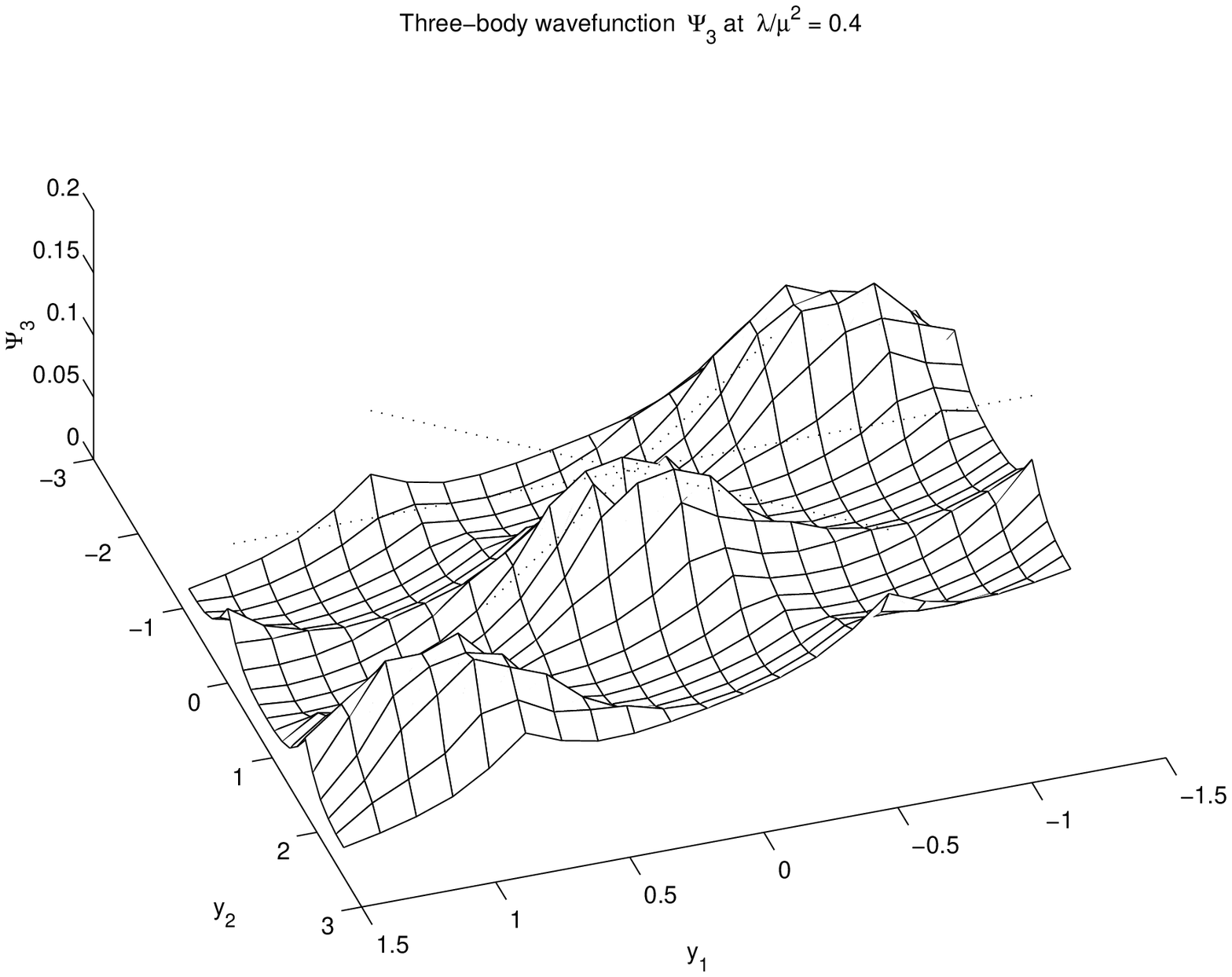}
\end{center}
\caption{Three-body wavefunction $\Psi_{3}$ at $\lambda=0.4\mu^{2}.$}%
\end{figure}\begin{figure}[ttttttttt]
\begin{center}
\epsfxsize=23pc \epsfbox{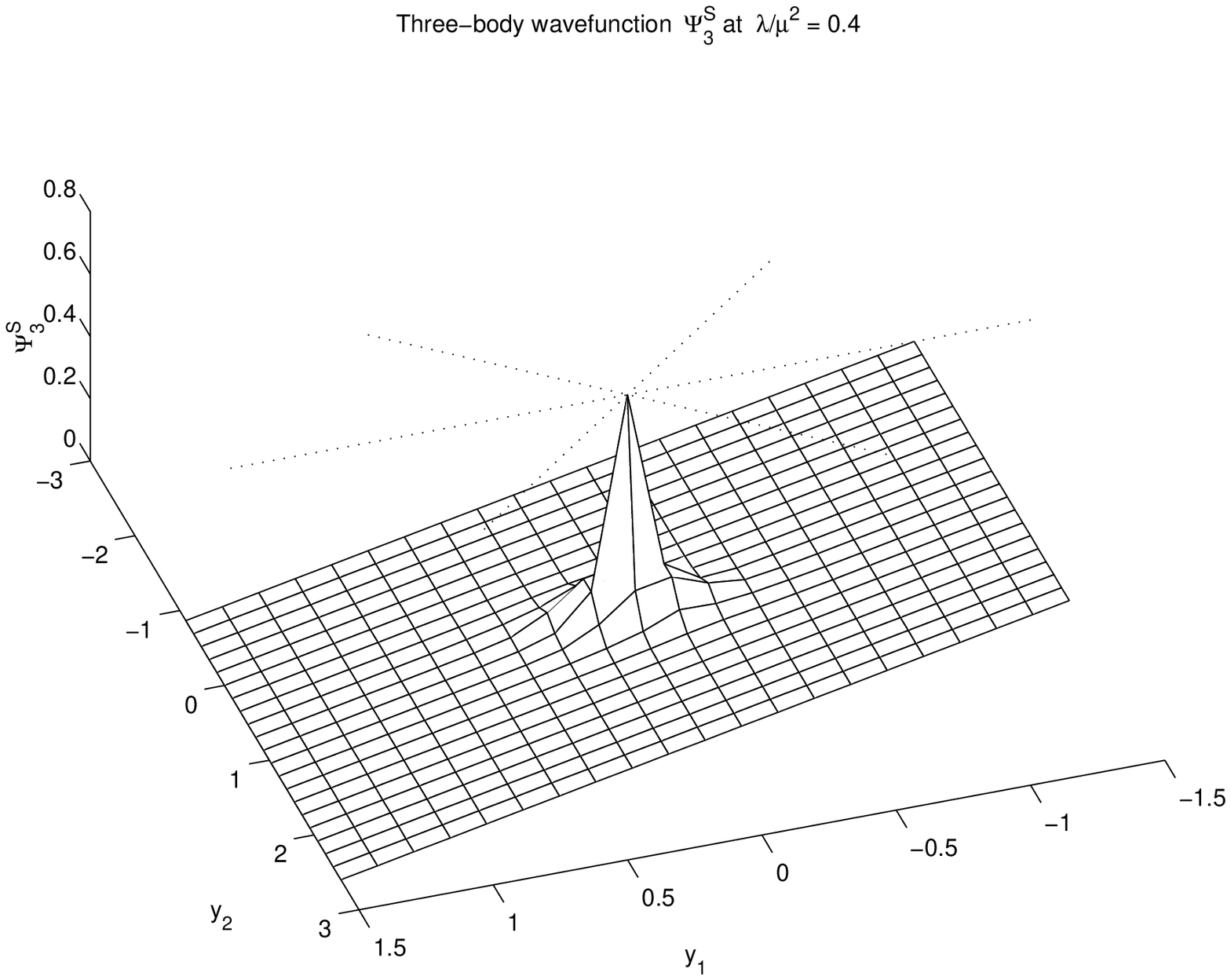}
\end{center}
\caption{Approximate three-body wavefunction $\Psi_{3}^{\text{S}}$ at
$\lambda=0.4\mu^{2}.$}%
\end{figure}\begin{figure}[tttttttttt]
\begin{center}
\epsfxsize=23pc \epsfbox{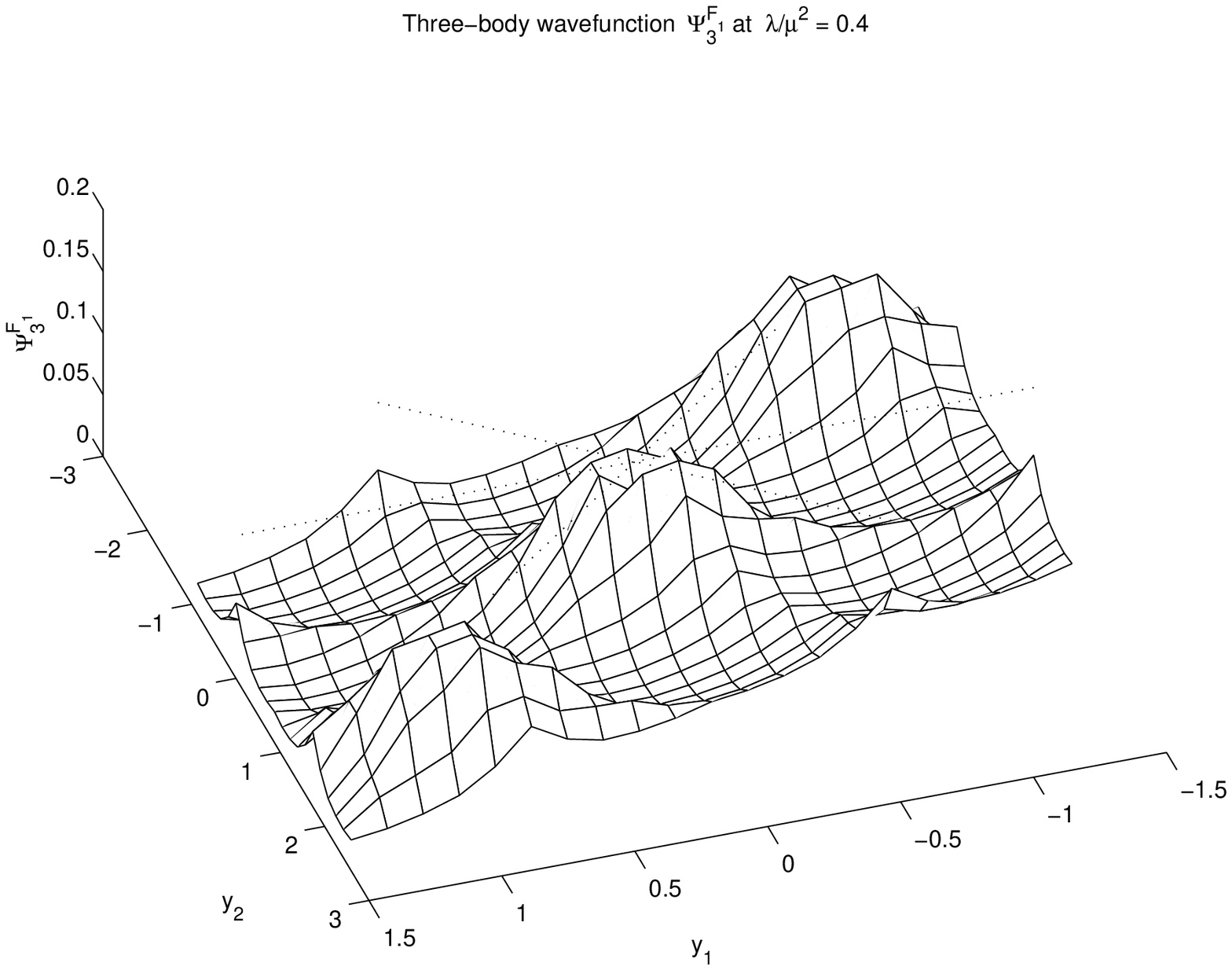}
\end{center}
\caption{Approximate three-body wavefunction $\Psi_{3}^{\text{F}_{1}}$ at
$\lambda=0.4\mu^{2}.$}%
\end{figure}\begin{figure}[ttttttttttt]
\begin{center}
\epsfxsize=23pc \epsfbox{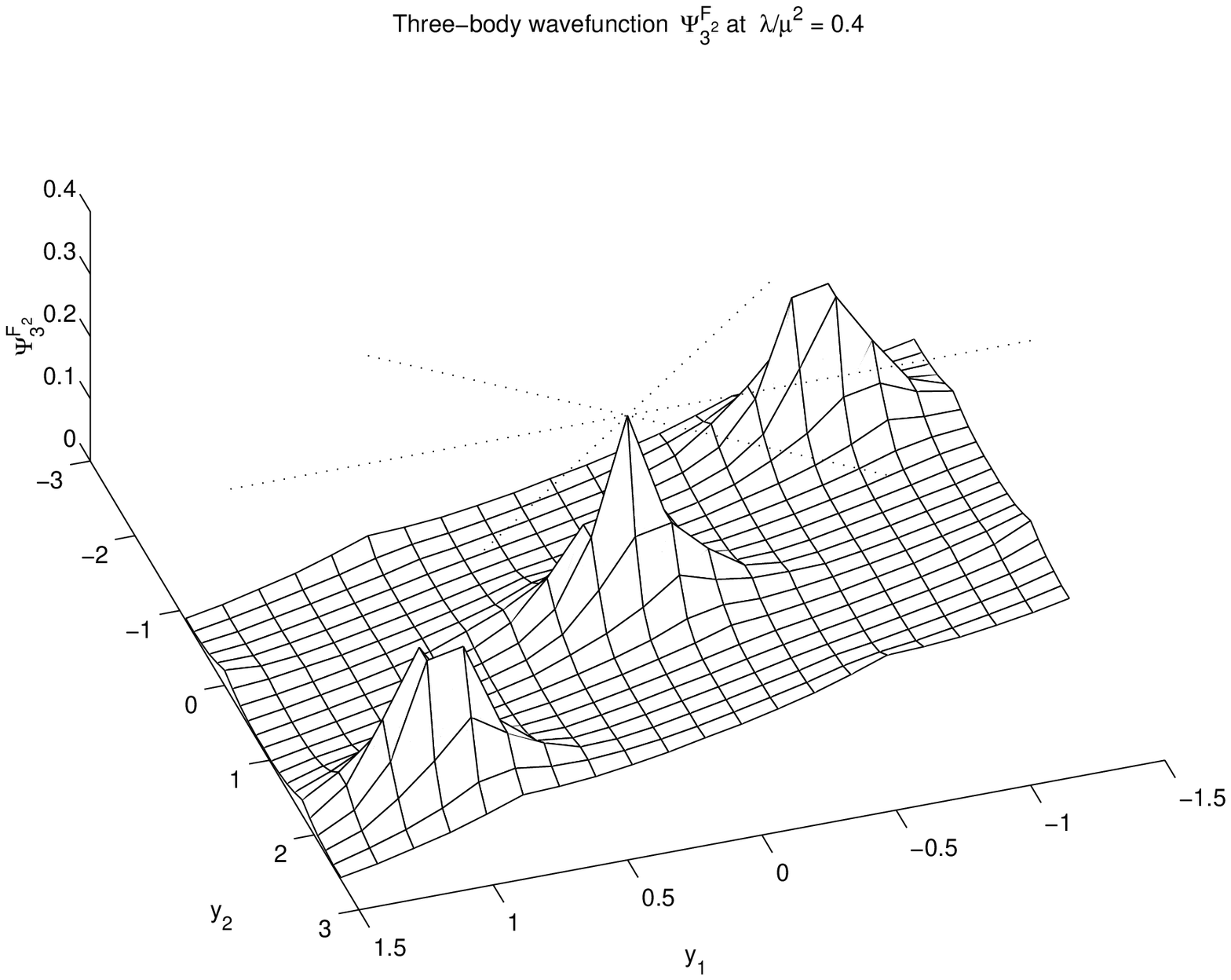}
\end{center}
\caption{Approximate three-body wavefunction $\Psi_{3}^{\text{F}_{2}}$ at
$\lambda=0.4\mu^{2}.$}%
\end{figure}
\end{document}